# Highly efficient light-emitting diodes based on intramolecular rotation

Dawei Di[1*], Alexander S. Romanov[2*], Le Yang[1*], Saul Jones[1], Richard H. Friend[1], Mikko Linnolahti[3†], Manfred Bochmann[2†], Dan Credgington[1†]

[1] Cavendish Laboratory, University of Cambridge, JJ Thomson Avenue, Cambridge, CB3 0HE, United Kingdom
[2] School of Chemistry, University of East Anglia, Earlham Road, Norwich, NR4 7TJ, United Kingdom
[3] Department of Chemistry, University of Eastern Finland, Joensuu Campus, FI-80101 Joensuu, Finland

[*] These authors contributed equally to this work.

[†]Correspondence to:  djnc3@cam.ac.uk, m.bochmann@uea.ac.uk, or mikko.linnolahti@uef.fi.

**Abstract**: The efficiency of an organic light-emitting diode (OLED) is fundamentally governed by the spin of recombining electron-hole pairs (singlet and triplet excitons), since triplets cannot usually emit light. The singlet-triplet energy gap, a key factor for efficient utilization of triplets, is normally positive. Here we show that in a family of materials with amide donor and carbene acceptor moieties linked by a metal, this energy gap for singlet and triplet excitons with charge-transfer character can be tuned from positive to negative values via the rotation of donor and acceptor about the metal-amide bond. When the gap is close to zero, facile intersystem crossing is possible, enabling efficient emission from singlet excitons. We demonstrate solution-processed LEDs with exceptionally high quantum efficiencies (near-100% internal and >27% external quantum efficiencies), and current and power efficiencies (87 cd A$^{-1}$ and 75 lm W$^{-1}$) comparable to, or exceeding, those of state-of-the-art vacuum-processed OLEDs and quantum dot LEDs.

**Main Text:** Since the demonstration of thin-film organic light-emitting diodes (OLEDs) in the late 1980s (*1, 2*), these devices have evolved from lab curiosity to a global industry encompassing smartphones, tablet computers, televisions and large-area lighting. The efficiency of an OLED is fundamentally governed by the spin of the tightly-bound electron-hole pairs (excitons) which form upon electrical excitation. The ratio of emissive singlet to dark triplet excitons formed from randomly spin-polarized electron and hole currents is 1:3. This sets an upper limit of 25% internal quantum efficiency (IQE) for fluorescent OLEDs since the exchange energy (the singlet-triplet energy gap) is generally large, around 0.5 eV, meaning that triplet excitons cannot intersystem cross to emissive singlets. Phosphorescent OLEDs, which utilize the heavy-atom effect in platinum or iridium compounds to render triplets emissive (*3, 4*), and thermally activated delayed fluorescence (TADF) OLEDs based on thermally-promoted triplet-to-singlet up-conversion for systems with low exchange energies (*5*), enable higher efficiencies. Here we show that the effective exchange energy in a family of copper and gold carbene metal amide compounds can be tuned via rotation about the metal-amide bond from positive to negative values. Importantly, when the effective exchange energy is close to zero, this enables facile intersystem crossing (ISC), which allows extremely efficient electroluminescence. These compounds are well-suited to solution-processing, which is a simple and inexpensive approach to fabrication compatible with large-area, roll-to-roll deposition. However, the precise control of device structure offered by vacuum-deposition usually enables significantly better performance. Herein, we report solution-processed LEDs with near-100% IQE (external quantum efficiencies >27%), high current and power efficiencies (87.1 cd $A^{-1}$ and 75.1 lm $W^{-1}$), as well as high brightness (up to $6.5 \times 10^4$ cd $m^{-2}$). These metrics are comparable to, or exceed, those of the state-of-the-art OLEDs (*5, 6*), and quantum dot LEDs (*7*). At brightness suitable for display applications, our best devices achieve the highest quantum efficiencies reported for any solution-processed LEDs without enhanced optical outcoupling.

Molecular rotation has attracted significant attention for its importance in biological and molecular nanomachines (*8*), and can be controlled by introducing chemical (*9*), thermal (*10, 11*), light (*10, 12*), and electrical (*12*) energy. Herein, we show that rotational flexibility can also be used to break the geometric equivalence of singlet and triplet excitons in OLEDs, by allowing each spin-state to relax into a different rotated geometry. All molecular emitters undergo some geometric reorganization in response to both charged and neutral excitations, leading to the well-known behavior of polaronic charge carriers and large Stokes shift. For singlet states with strong charge-transfer (CT) character, rotation that decouples the electron donor and acceptor (typically away from co-planarity) reduces the exchange interaction arising from overlapping electronic wavefunctions. Such reorganization leads to red-shifted fluorescence that accompanies the twisting of the intramolecular CT excited state (*13, 14*). As we show below from both modeling and experiment, the equivalent spin triplet CT state may behave oppositely, favoring the lower energy planar geometry, since there is no benefit to lowering the exchange interaction. Detailed calculations show that when singlet and triplet excitons undergo further relaxation within these different geometries, the relaxed singlet may lie lower in energy than the relaxed triplet. This inversion of the normal ordering of spin-state energies leads to efficient *down-conversion* from triplet excitons to lower-lying singlets, since a rotation angle must exist where the two states are degenerate. We term this process "rotationally accessed spin-state inversion" (RASI). In contrast to TADF systems, the degeneracy of the singlet and triplet levels is reached for a geometry which still gives strong oscillator strength for the singlet CT-character exciton (absorption coefficient above $10^4$ $cm^{-1}$). This facilitates prompt singlet emission (tens of ns) following ISC, and extremely high efficiency LED operation. For the systems reported here, the linkage between donor and acceptor is formed by a carbene-metal-amide bridge. This confers three important attributes: (i) extended π-electron delocalization, (ii) enhanced spin-orbit coupling, here from Cu and Au, and (iii) easy rotation of donor with respect to acceptor. Fig. 1A shows a schematic of the RASI process.

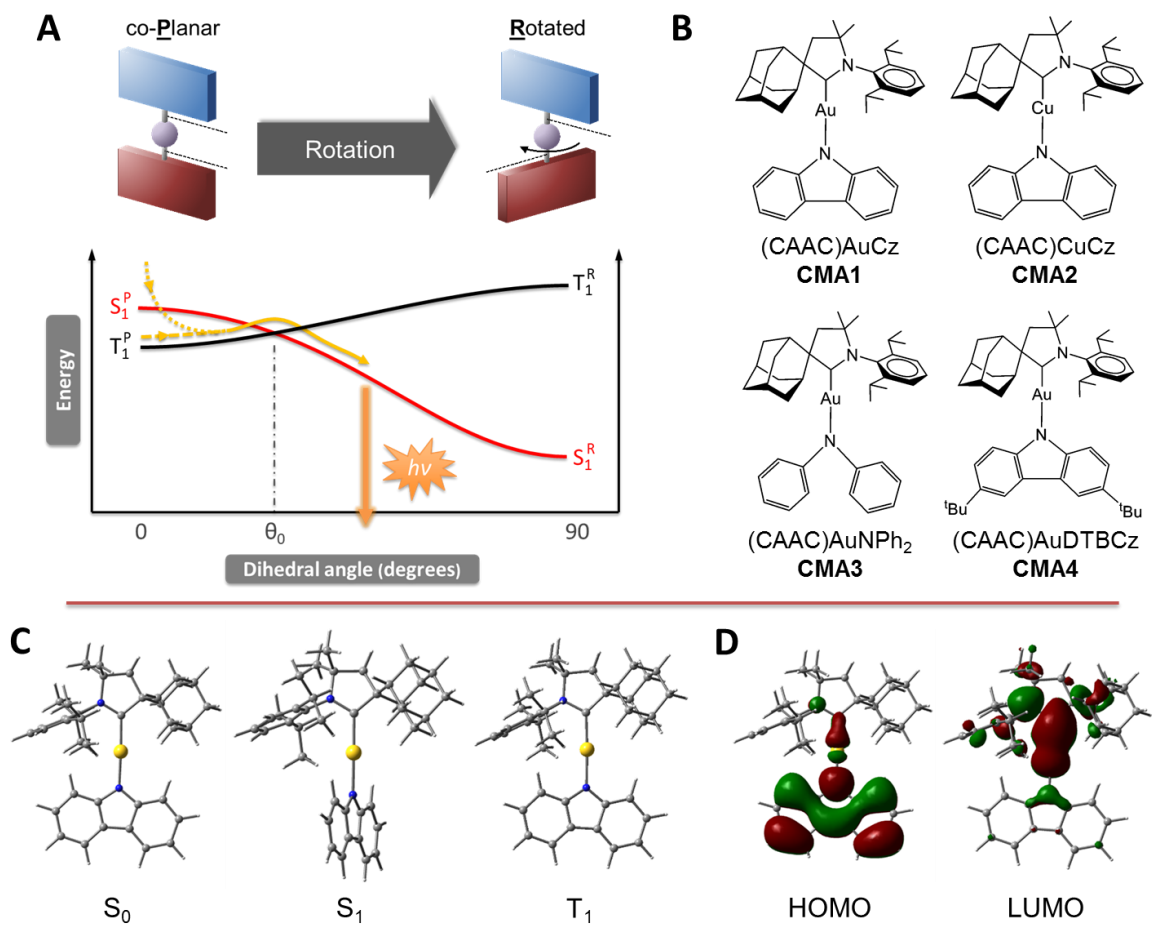

**Fig. 1** Rotationally accessed spin-state inversion mechanism, chemical structures and DFT calculations. **(A)** The RASI mechanism, described using a molecule with a rotational degree of freedom between electron donating and accepting moieties. When the donor and acceptor are coplanar (P), it follows $E(S_1) \geq E(T_1)$. In a rotated geometry (R), $E(S_1)$ may relax sufficiently that inversion of spin-state energies ($E(T_1) > E(S_1)$) can be achieved, with degeneracy occurring at $\theta_0$. Emission trajectories for photogenerated singlets (dot) and electrically generated triplets (dash) are indicated. **(B)** (CAAC)AuCz (CMA1) and its analogues. **(C)** Optimized molecular geometries of CMA1 for its ground state ($S_0$), first excited singlet ($S_1$) and triplet ($T_1$) states. **(D)** HOMO and LUMO of CMA1, obtained from DFT and TD-DFT calculations.

We designed and synthesized 2-coordinate cyclic (alkyl)(amino)-carbene (CAAC) compounds based on CAAC-M-amide (CMA) where M is Cu or Au. The compounds (CAAC)AuCz (CMA1) and its structurally similar analogues, (CAAC)CuCz (CMA2), (CAAC)AuNPh$_2$ (CMA3), and (CAAC)AuDTBCz (CMA4) are shown in Fig. 1B (Cz = carbazole anion, DTBCz = 3,6-di-tert.-butylcarbazole anion). These complexes are soluble in a range of organic solvents, do not undergo ligand rearrangement reactions in solution, and are thermally stable to > 270 °C (Table S1).

The ground ($S_0$), first excited singlet ($S_1$) and first excited triplet ($T_1$) states for CMA1-4 were computed using density functional theory (DFT) and time-dependent density functional theory (TD-DFT) with optimization of the molecular geometry. The ground state calculations yield close agreement with the X-ray crystal structures (Fig. S1). For all compounds, $S_0$ and the relaxed $T_1$ state correspond to a geometry with co-planar carbene and amide ligands, denoted by superscript $P$, while in the relaxed $S_1$ state the amide ligand is rotated by 90°, denoted by superscript $R$. (Figs. 1C and S2A). Both $S_1$ and $T_1$ states have strong CT character. Excitation from the highest-occupied molecular orbital (HOMO) to the lowest-unoccupied molecular orbital (LUMO) (Fig. 1D and Fig. S2B) contributes more than 70% of the $S_0 \rightarrow S_1$ transition. The significant overlap between HOMO and LUMO enabled by extended conjugation through the metal linkage means this transition is strongly allowed. We estimate its energy by considering the excited singlet constrained to the ground state geometry ($S_1^{P*}$). For CMA1, this yields a $S_0 \rightarrow S_1^{P*}$ transition at ~ 2.65 eV, in agreement with the measured absorption onset at ~450 nm (Fig. S3A) with high absorption coefficient (~$10^4$ cm$^{-1}$). The metal linkage also enables facile rotation about the metal-amide bond: the energy for full rotation in the $S_0$ ground state is calculated to be 143 meV for CMA1, equivalent to that of a freely rotatable carbon-carbon single bond (*15*).

Considering the relative state energies (Table S2), the relaxed co-planar triplet $T_1^P$ typically lies lower in energy than the equivalent singlet, $S_1^P$, for which the carbene-amide dihedral angle is fixed at 0°. However, increasing the dihedral angle allows $S_1$ to stabilize relative to $T_1$. Upon full relaxation, $S_1^R$ is lower than $T_1^P$, leading to a *negative* $S_1^R$-$T_1^P$ energy gap ($\Delta E_{ST}$) of -0.24 to -0.34 eV. Calculated $S_1$ and $T_1$ energies for dihedral angles between these extrema for CMA1 are presented in Fig. S4. These CMA molecules are therefore the first type of compound capable of realizing the rotationally accessed spin-state inversion pathway described in Fig. 1A.

The unusual energetic ordering in CMAs contrasts with carbene metal halide (CMH) fluorophores (*16*), where rotation about the metal-halide bond cannot lead to conformational change. For CMH compounds, calculated $\Delta E_{ST}$ (300-400 meV) was too large to allow TADF, and no phosphorescence was observed despite the presence of Au or Cu metal centers. The behavior of these linear CMA compounds also contrasts with 3-coordinate Cu(I) compounds where rotational freedom does not lead to spin-state inversion (*17*).

Solid films of CMA1-4 are photoluminescent, due to the lack of strong intermolecular interactions, as previously observed in CMHs (*16*). To understand the emission mechanism of this class of compounds, we investigated the photoluminescence (PL) spectra and kinetics of spin-coated thin films under pulsed laser excitation. All compounds show a fast component to the PL decay with lifetime less than 1 ns (see Figs. S3B & S3C) and a slow component whose lifetime is strongly dependent on temperature (vide infra). We focus below on detailed analysis of CMA1.

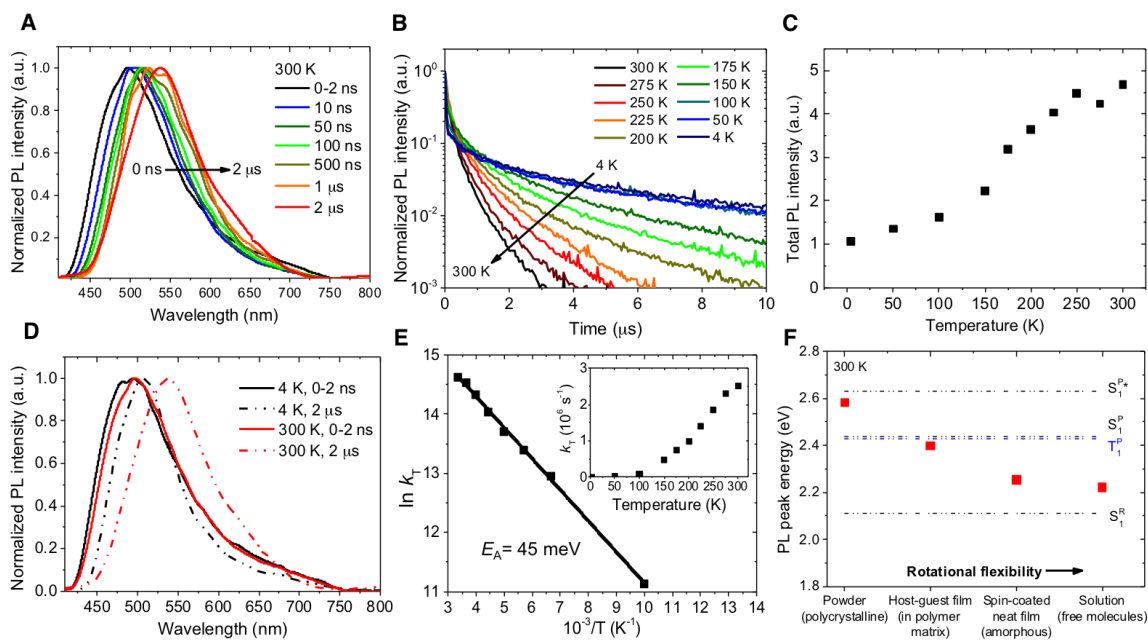

**Fig. 2** Photophysical characterization and temperature-dependent emission kinetics of CMA1. **(A)** Evolution of PL spectra with time at 300K. **(B)** Temperature-dependent PL kinetics. **(C)** Temperature-dependent total PL intensities calculated by integrating the time-dependent decays. **(D)** Fast (t < 2 ns) and slow (t = 2 µs) PL spectra recorded at 300K and 4K. **(E)** Temperature-dependent decay rate $k_T$, showing an activation energy of 45 meV above 100K. **(F)** Steady-state PL peak energies of CMA1 for different phases. Dashed lines show DFT calculations for selected excited state energies. The host-guest film comprises 20 wt% of CMA1 in PVK host. The solution sample is 0.1 wt% of CMA1 in chlorobenzene.

Time-resolved PL spectra of CMA1 on ns-µs timescales at 300K, Fig. 2A, show a red shift of the PL peak position from ~500 nm at early times (0-2 ns), to ~540 nm at 2 µs. The lifetime of the red-shifted slow component decreases from ~10 µs at 4K to ~350 ns at 300K (Fig. 2B). The total PL intensity increases significantly with temperature (Fig. 2C).

At 300K, relaxation of the initially photo-excited state through twisting and stretching modes is expected to occur on picosecond timescales. Rotation of the bulky amide group to a fully relaxed geometry occurs more slowly. Considering Fig. 2D, at 300K we assign the fast emission at ~ 2.5 eV to transitions from $S_1^P$ to $S_0$, following Kasha's rule.

At 4K, internal conversion to $S_1^P$ is slowed as relaxation modes are frozen out, slightly blue-shifting the fast emission. The lifetime of the fast emission is approximately independent of temperature (Fig. S3C) and its relative contribution is small (~5% of the time-integrated PL). Its fast decay rate ( > $10^9$ s$^{-1}$) indicates high oscillator strength, as expected for the co-planar $S_1^P \rightarrow S_0$ transition.

The slow emission exhibits two distinct regimes. Below 100K, it peaks at ~ 2.4 eV with lifetime of ~10 µs, independent of temperature. We assign this to weak $T_1^P \rightarrow S_0$ phosphorescence, enabled by metal-assisted spin-orbit coupling. The fast luminescence from the planar singlet manifold must therefore compete with ISC to $T_1^P$, the lowest energy state in the planar geometry. This is consistent with reversible quenching of the slow photoluminescence by exposure to oxygen. At higher temperature, these long-lived excitations can explore the lower-energy states associated with amide rotation, passing over the rotational barrier (at θ$_0$) where $S_1$ and $T_1$ are degenerate and crossing to the rotated singlet manifold (Fig. 1A). We therefore assign the slow red-shifted emission to the continuum of available $S_1 \rightarrow S_0$ transitions, and note that the luminescence efficiency from these states is approximately 5 times greater than emission from $T_1^P$ at low temperature (Fig. 2C). From the PL data presented in Fig. 2B, we calculate the activation energy ($E_A$) for the slow emission in the spin-coated CMA1 film to be 45 meV (Fig. 2E). This is considerably lower than the calculated energy for complete metal-amide bond rotation, and is accessible at room temperature. Calculated activation energies for CMA2-4 are very similar, and are shown in Fig. S5. A simple physical model for $E_A$ is outlined in the Supplementary Materials.

The observed emission behavior may be understood by considering the interplay between oscillator strength, rotation angle and the free volume available to each chromophore. Fig. 2F presents the steady-state PL spectra of CMA1 in phases with increasing free volume: from crystalline powder in which rotation is rigidly constrained, to solution phase in which molecules have considerable rotational freedom. Decreasing PL energy is associated with increasing free volume, in agreement with previous observations of solid-state molecular rotation (*18*). PL from crystalline powders is in excellent agreement with the fast emission of "frozen" spin-coated films at 4K (~ 2.6 eV), consistent with emission from constrained states where a limited range of relaxation modes are available. DFT calculations place $S_1$ in the crystalline geometry (Fig. S1) at 2.67 eV, similar to $S_1^{P*}$. In solution, the peak PL energy (2.2 eV) is consistent with emission from a fully relaxed $S_1$ population weighted by oscillator strength (Fig. S4) with a Boltzmann occupancy factor. The emission of amorphous CMA1 and CMA1 in a compact poly(9-vinylcarbazole) (PVK) matrix lie between the two, indicating that rotational relaxation is hindered, but not prevented, by neighboring molecules in these phases.

To examine the performance of an intramolecular rotation-based OLED (ROLED), we developed a simple fabrication route with all of the four organic layers fully solution-processed (Fig. 3A). The device energetic structure is shown in Fig. 3B. HOMO/LUMO energies of CMA1-4 measured using cyclic voltammetry are given in Table S3. A polymeric hole-transport/electron-blocking layer, poly(9,9-dioctylfluorene-co-N-(4-butylphenyl)diphenylamine) (TFB), was deposited on PEDOT:PSS without additional photo- or chemical crosslinking procedures (*19*), while CMA1-4 dispersed as a guest emitter in a PVK polymer host (*4, 20*) were used as the emissive layer, followed by a solution-deposited bathophenanthroline (BPhen) electron-transport/hole-blocking layer. From thickness profilometry measurements, we found that sequential deposition of the four solution-processed layers did not cause measurable thickness reduction of any of the underlying layers, suggesting minimum intermixing of the multilayer interfaces. The electroluminescence (EL) spectra and the external quantum efficiency (EQE) of the best ROLEDs are shown in Fig. 3C and Table 1 (with additional data in Figs. S6A & S6B. Angular emission profiles for the devices showed Lambertian emission (Fig. S6C), as is typical for OLEDs without microcavity outcoupling (*21*), which allows accurate estimation of EQE from on-axis irradiance. Similarly, the Commission Internationale de l'Éclairage (CIE) color coordinates of the devices showed no variation with EQE (Figs. S6D & S6E). Fig. 3D shows the maximum EQE histogram of 135 ROLEDs using CMA4, which produced the most efficient devices. Performance metrics are summarized in Table 1. The EQEs of the best devices at practical brightness (100 & 1000 cd m$^{-2}$) are higher than 25%, representing a record for solution-processed LEDs (*7, 22*) without enhanced optical outcoupling, while the peak EQE of our best device reached 27.5%. Using the same fabrication techniques, we prepared solution-processed OLEDs based on the high performance phosphorescent emitter Ir(ppy)$_3$ (*23*). The peak EQE achieved was ~18% (Fig. S6F), demonstrating the versatility of our fabrication method.

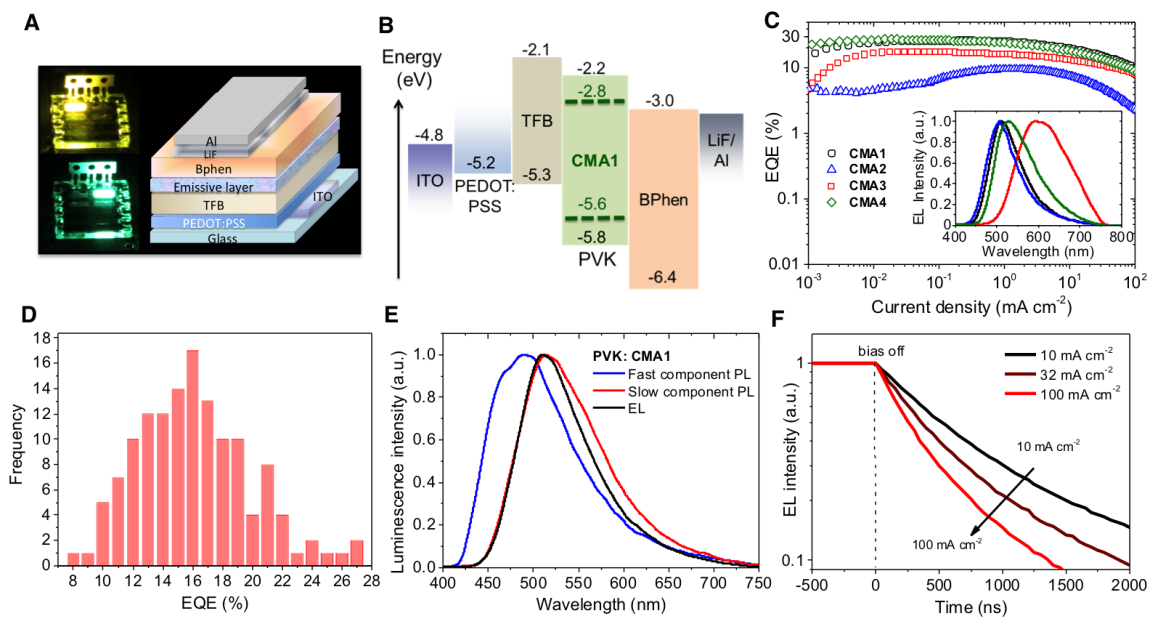

**Fig. 3:** Device performance and EL measurements. **(A)** Device structure of a solution-processed, multi-layer ROLED, with photographs of working devices. **(B)** Energy level diagram of a device based on CMA1. **(C)** EQE curves and EL spectra (inset). **(D)** Histogram of the maximum EQE measured from 135 devices based on CMA4. **(E)** Comparison of fast, slow PL and steady-state EL (for 20 wt% of CMA1 in PVK host). **(F)** Transient-EL curves measured after holding a device based on CMA1 at various steady-state current densities.

Assuming outcoupling efficiency in the 20-30% range expected for planar OLEDs (*5,7,21*), we infer a close-to-unity (80-100%) internal quantum efficiency for the best devices (based on CMA1&4). Time-resolved EL measurements confirm that 100% of emission occurs through the sub-µs slow emission channel (Figs. 3E & 3F). There is no fast (ns) component which would usually distinguish between geminate singlets and triplets (*24, 25*). EL spectra typically correspond to the lowest accessible excited state, since recombination will preferentially occur at these sites. The steady-state EL spectrum agrees very well with that of the lowest-energy PL (Fig. 3E). Together, these imply that all excitons, regardless of initial spin, contribute to luminescence in these materials, and that emission primarily occurs from the low-energy rotated singlet manifold identified from PL measurements and predicted by DFT calculation. The lifetime of the electroluminescence decreases with increasing current density, which may originate from exciton quenching processes (*26*). Given the relatively slow efficiency roll-off at high current, this may also be an effect of Joule heating increasing the emission rate. At present, it is difficult to separate these effects, but it is clear there is scope for improvement at high current densities.

**Table 1:** Summary of ROLED performances (best devices). Efficiency values at different brightness (100 cd m$^{-2}$ and 1000 cd m$^{-2}$) are shown.

| Emitter | Turn-on Voltage (V) | EQE (%) | Current Efficiency (cd A$^{-1}$) | Power Efficiency (lm W$^{-1}$) | Max. Luminance (cd m$^{-2}$) |
|---|---|---|---|---|---|
| | | Max/100/1000 cd m$^{-2}$ | Max/100/1000 Cd m$^{-2}$ | Max/100/1000 cd m$^{-2}$ | |
| **CMA1** | 2.6 | 26.3/ 26.1/ 25.2 | 76.3/ 75.8/ 73.0 | 62.7/ 50.0/ 37.0 | 44700 |
| **CMA2** | 3.4 | 9.7/ 8.9/ 9.2 | 30.4 /28.0/ 29.0 | 11.8/ 11.7/ 9.3 | 7790 |
| **CMA3** | 3.0 | 17.9/ 17.3/ 15.5 | 45.2/ 43.7/ 39.1 | 33.6/ 25.0/ 17.0 | 39540 |
| **CMA4** | 2.6 | 27.5/ 26.6/ 24.5 | 87.1/ 84.5/ 77.9 | 75.1/ 50.2/ 35.5 | 64700 |

As described above, we consider that the rapid ISC from triplet to singlet required for this efficient device performance is achieved at rotational configurations where the singlet and triplet energies are approximately degenerate. This allows spin-orbit coupling, enhanced by the involvement of the metal atom in the transition orbitals, to mix the two spin states effectively (Fig. 1A). For CMA1, calculations suggest this occurs at low dihedral angles, with $\theta_0$ around 15° (Fig S4). While we cannot rule out a contribution from activated exciton migration, the low activation energies we measure likely correspond to the partial rotation required to achieve ISC; further rotation in the singlet manifold is energetically downhill, hindered only by the surrounding matrix. The emission rate is thus limited primarily by rotationally-accessed ISC, rather than the radiative rate; even the lowest energy emission from spin-cast CMA1 film occurs from singlet states with intrinsic radiative lifetimes estimated to be in the tens of nanoseconds range. This suggests the use of host materials with greater free volume for guest rotation to achieve the highest possible emission rates.

In conclusion, we demonstrate high-performance solution-processed OLEDs utilizing a new class of material to harvest the energy of normally non-emissive triplet excitons. Our work closes the performance gap between solution-processed and vacuum-deposited OLEDs. We take a novel approach to the design of molecular photoemitters, employing a linear carbene-metal-amide linkage between luminescent donor and acceptor moieties to simultaneously enhance spin-orbit coupling and oscillator strength while reducing the barrier to rotation between donor and acceptor. The result is a family of compounds in which the normal energetic ordering of spin states may be inverted via intramolecular rotation. Photoluminescence from these materials is characterized by both fast and slow emissive species, where the slow species red-shifts with increasing temperature and free volume, and the overall PL yield increases significantly with temperature. These contrast with the behavior of both TADF compounds (*5*) and iridium-based phosphorescent emitters (*27, 28*). Crucially, degenerate mixing of spin states occurs without sacrificing radiative lifetime, allowing fluorescence to compete effectively with non-radiative relaxation. At 300K, the emission lifetime for all compounds is well below 1 μs, considerably shorter than most organometallic TADF emitters (whose typical lifetimes are 5-10 μs (*29*)), offering a significant advantage for achieving high-brightness operation and stability (*26*). Beyond OLEDs, spin-state inversion opens a new route for the design of molecular optoelectronics. For example, spontaneous down-conversion from triplets to singlets is highly desirable for the realization of electrically-pumped organic lasers. The possibilities of modulating the rotational motion of the emissive molecules with thermal and electromagnetic energies could excite further interest in the development of optoelectronic nanomachines.

**References:**

1. C. Tang, S. VanSlyke, Organic electroluminescent diodes. *Appl. Phys. Lett*. **51**, 913 (1987).
2. J. Burroughes, D. Bradley, A. Brown, R. Marks, K. Mackay, R. Friend, P. Burns, A. Holmes, Light-emitting diodes based on conjugated polymers, *Nature* **347**, 539-541 (1990).
3. S. Forrest, M. Baldo, D. O'Brien, Y. You, A. Shoustikov, S. Sibley, M. Thompson, Highly efficient phosphorescent emission from organic electroluminescent devices, *Nature* **395**, 151-154 (1998).
4. Y. Ma, H. Zhang, J. Shen, C. Che, Electroluminescence from triplet metal—ligand charge-transfer excited state of transition metal complexes, *Synthetic Metals* **94**, 245-248 (1998).
5. H. Uoyama, K. Goushi, K. Shizu, H. Nomura, C. Adachi, Highly efficient organic light-emitting diodes from delayed fluorescence, *Nature* **492**, 234-238 (2012).
6. S. Reineke, F. Lindner, G. Schwartz, N. Seidler, K. Walzer, B. Lüssem, K. Leo, White organic light-emitting diodes with fluorescent tube efficiency, *Nature* **459**, 234-238 (2009).
7. X. Dai, Z. Zhang, Y. Jin, Y. Niu, H. Cao, X. Liang, L. Chen, J. Wang, X. Peng, Solution-processed, high-performance light-emitting diodes based on quantum dots, *Nature* **515**, 96-99 (2014).
8. W. Browne, B. Feringa, Making molecular machines work, *Nature Nanotech* **1**, 25-35 (2006).
9. S. Fletcher, F. Dumur, M. Pollard, B. Feringa, A reversible, unidirectional molecular rotary motor driven by chemical energy, *Science* **310**, 80-82 (2005).
10. N. Koumura, R. Zijlstra, R. van Delden, N. Harada, B. Feringa, B. Light-driven molecular rotor. *Nature* **401**, 152–155 (1999).
11. D. Leigh, J. Wong, F. Dehez, F. Zerbetto, Unidirectional rotation in a mechanically interlocked molecular rotor. *Nature* **424**, 174–179 (2003).
12. M. Hawthorne, J. Zink, J. Skelton, M. Bayer, C. Liu, E, Livshits, R. Baer, D. Neuhauser, Electrical or Photocontrol of the Rotary Motion of a Metallacarborane, *Science* **303**, 1849-1851 (2004).
13. S. Sasaki, G. Drummen, G. Konishi, Recent advances in twisted intramolecular charge transfer (TICT) fluorescence and related phenomena in materials chemistry, *J. Mater. Chem. C* **4**, 2731-2743 (2016).
14. A. Siemiarczuk, Z. Grabowski, A. Krówczyński, M. Asher, M. Ottolenghi, Two emitting states of excited p-(9-anthryl)-n,n-dimethylaniline derivatives in polar solvents, *Chemical Physics Letters* **51**, 315-320 (1977).
15. J. Zheng, K. Kwak, J. Xie, M. Fayer, Ultrafast Carbon-Carbon Single-Bond Rotational Isomerization in Room-Temperature Solution, *Science* **313**, 1951-1955 (2006).

16. A. Romanov, D. Di, L. Yang, J. Fernandez-Cestau, C. Becker, C. James, B. Zhu, M. Linnolahti, D. Credgington, M. Bochmann, Highly photoluminescent copper carbene complexes based on prompt rather than delayed fluorescence, *Chem. Commun.* 52, 6379-6382 (2016).
17. M. Leitl, V. Krylova, P. Djurovich, M. Thompson, H. Yersin, Phosphorescence versus thermally activated delayed fluorescence. Controlling singlet-triplet splitting in brightly emitting and sublimable Cu(I) compounds. *J. Am. Chem. Soc.* **45**, 16032–16038 (2014).
18. K. Al-Hassan, T. Azumi, The role of free volume in the twisted intramolecular charge transfer (TICT) emission of dimethylaminobenzonitrile and related compounds in rigid polymer matrices, *Chemical Physics Letters* **146**, 121-124 (1988).
19. R. Png, P. Chia, J. Tang, B. Liu, S. Sivaramakrishnan, M. Zhou, S. Khong, H. Chan, J. Burroughes, L. Chua, R. Friend, P. Ho, High-performance polymer semiconducting heterostructure devices by nitrene-mediated photocrosslinking of alkyl side chains, *Nature Materials* **9**, 152-158 (2009).
20. D. Abbaszadeh, A. Kunz, G. Wetzelaer, J. Michels, N. Crǎciun, K. Koynov, I. Lieberwirth, P. Blom, Elimination of charge carrier trapping in diluted semiconductors, *Nature Materials* **15**, 628-633 (2016).
21. N. Greenham, R. Friend, D. Bradley, Angular Dependence of the Emission from a Conjugated Polymer Light-Emitting Diode: Implications for efficiency calculations, *Adv. Mater.* **6**, 491-494 (1994).
22. N. Aizawa, Y. Pu, M. Watanabe, T. Chiba, K. Ideta, N. Toyota, M. Igarashi, Y. Suzuri, H. Sasabe, J. Kido, Solution-processed multilayer small-molecule light-emitting devices with high-efficiency white-light emission, *Nature Communications* **5**, 5756 (2014).
23. M. Baldo, S. Lamansky, P. Burrows, M. Thompson, S. Forrest, Very high-efficiency green organic light-emitting devices based on electrophosphorescence, *Appl. Phys. Lett.* **75**, 4 (1999).
24. D. Kondakov, Characterization of triplet-triplet annihilation in organic light-emitting diodes based on anthracene derivatives, *J. Appl. Phys.* **102**, 114504 (2007).
25. B. Wallikewitz, D. Kabra, S. Gélinas, R. Friend, Triplet dynamics in fluorescent polymer light-emitting diodes, *Phys. Rev. B* **85**, 045209 (2012).
26. N. Giebink, S. Forrest, Quantum efficiency roll-off at high brightness in fluorescent and phosphorescent organic light emitting diodes, *Phys. Rev. B* **77**, 235215 (2008).
27. T. Sajoto, P. Djurovich, A. Tamayo, J. Oxgaard, W. Goddard, M. Thompson, Temperature Dependence of Blue Phosphorescent Cyclometalated Ir(III) Complexes, *J. Am. Chem. Soc.* **131**, 9813-9822 (2009).
28. K. Goushi, Y. Kawamura, H. Sasabe, C. Adachi, Unusual Phosphorescence Characteristics of Ir(ppy)$_3$ in a Solid Matrix at Low Temperatures, *Jpn. J. Appl. Phys.* **43**, L937-L939 (2004).
29. Y. Tao, K. Yuan, T. Chen, P. Xu, H. Li, R. Chen, C. Zheng, L. Zhang, W. Huang, Thermally Activated Delayed Fluorescence Materials Towards the Breakthrough of Organoelectronics, *Adv. Mater.* **26**, 7931-7958 (2014).
30. A. Romanov, M. Bochmann, Gold(I) and Gold(III) Complexes of Cyclic (Alkyl)(amino)carbenes, *Organometallics* **34**, 2439-2454 (2015).
31. G. Gritzner, J. Kůta, Recommendations on reporting electrode potentials in nonaqueous solvents, *Electrochimica Acta* **29**, 869-873 (1984).
32. *Programs CrysAlisPro*, Oxford Diffraction Ltd., Abingdon, UK (2010).

33. G. Sheldrick, A short history of SHELX, *Acta Cryst Sect A* **64**, 112-122 (2007).
34. F. Furche, D. Rappoport, Density functional methods for excited states: equilibrium structure and electronic spectra in *Computational Photochemistry* (ed. M. Olivuccim) 93-128 (Elsevier, Amsterdam, 2005).
35. J. Perdew, K. Burke, M. Ernzerhof, Generalized Gradient Approximation Made Simple, *Phys. Rev. Lett.* **77**, 3865-3868 (1996).
36. C. AdamoV. Barone, Toward reliable density functional methods without adjustable parameters: The PBE0 model, *The Journal of Chemical Physics* **110**, 6158 (1999).
37. F. Weigend, M. Häser, H. Patzelt, R. Ahlrichs, RI-MP2: optimized auxiliary basis sets and demonstration of efficiency, *Chemical Physics Letters* **294**, 143-152 (1998).
38. F. Weigend, R. Ahlrichs, Balanced basis sets of split valence, triple zeta valence and quadruple zeta valence quality for H to Rn: Design and assessment of accuracy, *Phys. Chem. Chem. Phys.* **7**, 3297 (2005).
39. K. Peterson, D. Figgen, E. Goll, H. Stoll, M. Dolg, Systematically convergent basis sets with relativistic pseudopotentials. II. Small-core pseudopotentials and correlation consistent basis sets for the post-d group 16–18 elements, *The Journal of Chemical Physics* **119**, 11113 (2003).
40. D. Andrae, U. Haeussermann, M. Dolg, H. Stoll, H. Preuss, Energy-adjustedab initio pseudopotentials for the second and third row transition elements, *Theoretica Chimica Acta* **77**, 123-141 (1990).
41. M. Bühl, C. Reimann, D. Pantazis, T. Bredow, F. Neese, Geometries of Third-Row Transition-Metal Complexes from Density-Functional Theory, *J. Chem. Theory Comput.* **4**, 1449-1459 (2008).
42. R. Kang, H. Chen, S. Shaik, J. Yao, Assessment of Theoretical Methods for Complexes of Gold(I) and Gold(III) with Unsaturated Aliphatic Hydrocarbon: Which Density Functional Should We Choose?, *J. Chem. Theory Comput.* **7**, 4002-4011 (2011).
43. C. Adamo, G. Scuseria, V. Barone, Accurate excitation energies from time-dependent density functional theory: Assessing the PBE0 model, *The Journal of Chemical Physics* **111**, 2889 (1999).
44. C. Adamo, V. Barone, R. Subra, The mechanism of spin polarization in aromatic free radicals, *Theoretical Chemistry Accounts: Theory, Computation, and Modeling (Theoretica Chimica Acta)* **104**, 207-209 (2000).
45. D. Jacquemin, A. Planchat, C. Adamo, B. Mennucci, TD-DFT Assessment of Functionals for Optical 0–0 Transitions in Solvated Dyes, *J. Chem. Theory Comput.* **8**, 2359-2372 (2012).
46. A. Laurent, M. Medved, D. Jacquemin, Using TD-DFT to probe the nature of donor-acceptor Stenhouse adduct (DASA) photochromes, *ChemPhysChem* , DOI: 10.1002/cphc.201600041 (2016).
47. I. Koshevoy, Y. Lin, A. Karttunen, P. Chou, P. Vainiotalo, S. Tunik, M. Haukka, T. Pakkanen, Intensely Luminescent Alkynyl−Phosphine Gold(I)−Copper(I) Complexes: Synthesis, Characterization, Photophysical, and Computational Studies, *Inorganic Chemistry* **48**, 2094-2102 (2009).
48. I. Koshevoy, Y. Chang, Y. Chen, A. Karttunen, E. Grachova, S. Tunik, J. Jänis, T. Pakkanen, P. Chou, Luminescent Gold(I) Alkynyl Clusters Stabilized by Flexible Diphosphine Ligands, *Organometallics* **33**, 2363-2371 (2014).
49. Y. Zhao, D. Truhlar, The M06 suite of density functionals for main group thermochemistry, thermochemical kinetics, noncovalent interactions, excited states, and transition elements:

two new functionals and systematic testing of four M06 functionals and 12 other functionals, *Theor Chem Account* **119**, 525-525 (2008).
50. M. Frisch, G. Trucks, H. Schlegel, G. Scuseria, M. Robb, J. Cheeseman, G. Scalmani, V. Barone, B. Mennucci, G. Petersson, H. Nakatsuji, M. Caricato, X. Li, H. Hratchian, A. Izmaylov, J. Bloino, G. Zheng, J. Sonnenberg, M. Hada, M. Ehara, K. Toyota, R. Fukuda, J. Hasegawa, M. Ishida, T. Nakajima, Y. Honda, O. Kitao, H. Nakai, T. Veven, J. A. Montgomery, Jr., J. Peralta, F. Ogliaro, M. Bearpark, J. Heyd, E. Brothers, K. Kudin, V. Staroverov, T. Keith, R. Kobayashi, J. Normand, K. Raghavachari, A. Rendell, J. Burant, S. Iyengar, J. Tomasi, M. Cossi, N. Rega, J. M. Millam, JM. Klene, J. E. Knox, J. B. Cross, V. Bakken, C. Adamo, J. Jaramillo, R. Gomperts, R. Stratmann, O. Yazyev, A. Austin, R. Cammi, C. Pomelli, J. Ochterski, R. Martin, K. Morokuma, V. Zakrzewski, G. Voth, P. Salvador, J. Dannenberg, S. Dapprich, A. Daniels, O. Farkas, J. Foresman, J. Ortiz, J. Cioslowski, D. Fox. Gaussian 09, Revision C.01, Gaussian, Inc., Wallingford, CT (2010).

**Acknowledgments:** We thank Dr. M. Roberts and Cambridge Display Technology Limited (Company number 02672530) for careful re-calibration of our OLED efficiency measurements and for the angular emission measurements. We thank H. Stern and Dr. A. Sadhanala for their help with the cryogenic measurements. We also thank J. Richter for his help and suggestions on the ICCD measurements. B. Zhu and L. Meraldi are acknowledged for their assistance in preliminary work. J. Nisbett is acknowledged for assistance with EL measurements. D.D. and R.H.F. acknowledge the Department of Physics (University of Cambridge) and the KACST–Cambridge University Joint Centre of Excellence for support. L.Y. thanks the Singapore Agency for Science, Technology and Research (A*STAR) for a PhD studentship. M.L. acknowledges support by the Academy of Finland (Project 251448). The computations were made possible by use of the Finnish Grid Infrastructure and Cloud Infrastructure resources. This work was supported by the European Research Council. M.B. is an ERC Advanced Investigator Award holder (grant no. 338944-GOCAT). D.C. and S.J. acknowledge the Royal Society for financial support.

**Author contributions:** D.D. and L.Y. developed and characterized the OLED devices. A.S.R. set up the collaboration and performed the molecular design, synthesis, X-ray crystallography and electrochemistry. D.D., L.Y. and S.J. carried out the photoluminescence studies. D.D. and L.Y. performed the electroluminescence experiments. D.C., M.B. and D.D. planned the study and designed the experiments. M.B. conceived the initial concept of the emission mechanism, which was further developed by D.D., D.C. and R.H.F. M.L. performed the quantum chemical calculations. D.D., D.C., M.B., L.Y. and R.H.F. co-wrote the manuscript.